\documentclass[10pt]{article}
\usepackage[utf8]{inputenc}
\usepackage{color}
\usepackage{verbatim}
\usepackage{cite}
\usepackage[colorlinks,linkcolor=blue,citecolor=green]{hyperref}

\usepackage{graphicx}
\usepackage{amsmath,amsfonts,amssymb,amsthm, bm}
\usepackage{nccmath}
\usepackage{enumerate}
\usepackage{multirow}
\usepackage{subfigure}
\usepackage{dsfont}
\usepackage{slashed}
\usepackage{textcomp}
\usepackage{url}
\usepackage{marvosym}
\usepackage{wasysym}
\usepackage{float}
\usepackage{capt-of}
\usepackage{subfigure} 
\usepackage{ragged2e}
\usepackage{xcolor}
\usepackage{authblk}

\allowdisplaybreaks[3]

\usepackage{mathrsfs}
\usepackage[scr=boondox]{mathalfa}

\graphicspath{{figures/}}

\author[1]{Pierandrea Vergallo}
\author[2]{Bennet Karetta}
\author[1,3]{Giancarlo Consolo}
\author[2]{Olena Gomonay \thanks{ogomonay@uni-mainz.de}}

\affil[1]{Istituto Nazionale di Fisica Nucleare – Sez. Lecce}
\affil[2]{Institute of Physics, Johannes Gutenberg-University Mainz, 55099 Mainz, Germany}
\affil[3]{Department of Mathematical, Computer,Physical and Earth Sciences University of Messina,V.le F. Stagno D'Alcontres 31, I-98166 Messina, Italy}
\date{ \today}
 
\begin{document}
\title{Domain-wall orientation in antiferromagnets controlled by magnetoelastic effects}

\maketitle
 

\vspace{0.5cm}
\begin{abstract}
In this paper, we develop the mathematical framework to describe the physical phenomenon behind the equilibrium configuration joining two antiferromagnetic domains. We firstly define the total energy of the system and deduce the governing equations by minimizing it with respect to the field variables. Then, we solve the resulting system of nonlinear PDEs together with proper initial and boundary conditions by varying the orientation of the 90$^{\circ}$ domain wall (DW) configuration along the sample. Finally, the angular dependence of elastic and magnetoelastic energies as well as of incompatibility-driven volume effects is computed. 
\end{abstract}

\vspace{0.5cm}


 \section{Introduction}
\label{sec::introduction}
Antiferromagnets are considered as prospective materials for spintronic applications because of their fast internal magnetic dynamics and robustness with respect to the external magnetic field. Those of antiferromagnets that are suitable for practical applications due to high ordering temperature and high, teraherz,  frequencies are also known for pronounced magnetoelastic coupling. Magnetoelastic effects are usually neglected while considering antiferromagnetic dynamics and switching. However, they can pin the domain walls \cite{Gomonay:PhysRevLett.91.237205}, modify magnon spectra \cite{Bossini:PhysRevLett.127.077202, Gomonay_2021, Bauer:PhysRevB.104.014403}, stabilise antiferromagnetic textures in the finite size samples \cite{Meer:PhysRevB.106.094430, Folven2010, Folven2012, Folven:10.1063/1.5116806, Gomonay2002a, Reimers2022}. As such, magnetoelastic effects play an important role in the dynamics of antiferromagnets and with the proper tailoring can even open new functionalities.

Recently we \cite{Wittmann:PhysRevB.106.224419, Meer:PhysRevB.106.094430} developed an approach for description of magnetoelastic effects in the finite-size antiferromagnets based on the concept of magnetoelastic charges and applied it to the formation of the domain structure in thin films. This approach was inspired by the works of Kleman  \cite{Kleman1972, Kleman1974a, kleman1978magnetostriction}, who suggested to use dislocation theory in magnetoelasticity of ferromagnets. The key idea is to split deformations of a magnetic material into two parts: spontaneous (plastic) strains $\epsilon^{sp}$ that are induced by the magnetic ordering in an uncontrained infinite sample, and additional, elastic strains. Spontaneous strains  are associated with distribution of the magnetic vectors. However, while the magnetic moments are continuously distributed in the magnetic textures, corresponding  spontaneous strains can be incompatible with each other and break continuity of the material. Additional elastic strains compensate incompatibility and restore continuity of the sample. Formally,  incompatibilities of spontaneous strains are treated as magnetoelastic charges, and corresponding elastic strains are similar to magnetostatic fields usually discussed in application to ferromagnetic materials. At the microscopic level the incompatibility charges are associated with the continuous distribution of dislocations. 

 Due to mathematical similarity between equations of magnetostatics and magnetoelasticity, magnetoelastic charges produce similar effects as dipole-dipole fields in ferromagnets. In particular, incompatibility of strains inside the domain wall can result in a pronounced increase of the domain wall energy and thus set preferable orientation of the domain wall. Strain-controlled orientation of the domain wall was discussed earlier in application to ferroleastics  \cite{Sapriel:PhysRevB.12.5128}, ferroelectrics \cite{Bishop2003}, and ferromagnets \cite{Kleman1972} based mainly on geometrical considerations and in assumption of atomically sharp domain walls. In this paper we extend study of the domain walls to antiferromagnets, in which magnetoelastic forces are the only sources of long-range interactions. As such, magnetoelastic effects in antiferromagnets are not covered up by demagnetizing or  depolarizing effects, which is the case in other ferroics (ferromagnets, ferroelectrics). Moreover, we remove the assumption of infinitely sharp domain wall and consider finite-size  samples, relevant for spintronic applications
 
 We calculate distribution of the elastic strains and corresponding contribution into the energy for different orientations of the 90$^\circ$ domain wall in a generic easy-plane antiferromagnet. We show that configurations with the minimal energy correspond to orientation of the domain wall along hard magnetic axes, in agreement with symmetry-based predictions. We demonstrate that angular-dependent contribution into the energy of the domain wall originates from the volume (isotropic) strains which opens a possibility to manipulate orientation of the domain walls using  thermal effects. 





\section{Mathematical model of a 90° Domain Wall}
\label{sec::math}
We consider a thin film of a collinear antiferromagnet with easy-plane anisotropy. The film thickness is $d$, the film plane coincides with the easy plane. The sample has a square shape with the side length  $2L\gg d$. We introduce coordinates axes $x$ and $y$ along the sample edges, $z$-axis is parallel to the film plane.  For description of the magnetic system we assume four-fold symmetry within the plane. As such, there are four  equivalent easy magnetic directions corresponding to equilibrium orientations of the N\'eel vector.  Due to strong out-of-plane anisotropy, the N\'eel vector stays within the film plane and can be parametrised with the only angle $\theta(x,y)$  calculated from easy direction: $\mathbf{n}=(\cos \theta, \sin \theta, 0)$.  In what follows we consider domain wall separating two states with orthogonal orientations of  the N\'eel vector.  The domain wall center coincides with the sample center $x=0$, $y=0$.

We use approximation of linear elasticity and describe the deformed state with the displacement vector  $\textbf{u}=(u,v,w)$ and corresponding  strain tensor $\epsilon$ a\begin{equation}
\epsilon=\begin{pmatrix}
\epsilon_{11}&\epsilon_{12}&\epsilon_{13}\\
\epsilon_{12}&\epsilon_{22}&\epsilon_{23}\\
\epsilon_{13}&\epsilon_{23}&\epsilon_{33}\\
\end{pmatrix}
=\frac{1}{2}\left(\nabla \textbf{u}+ \nabla \textbf{u}^\text{T}\right)=\displaystyle\begin{pmatrix}
u_{x}&\frac{u_{y}+v_{x}}{2}&\frac{u_z+w_x}{2}\medskip \\ \frac{u_{y}+v_{x}}{2}&v_{y}&\frac{v_z+w_y}{2}\medskip \\ \frac{u_z+w_x}{2}&\frac{v_z+w_y}{2}&w_z
\end{pmatrix}
\label{eq::math_epsi}
\end{equation}
In \eqref{eq::math_epsi}, subscripts denote partial derivative with respect to the indicated variable and $\text{T}$ stands for the transpose.

Distribution of the N\'eel vector and strains is calculated from minimization of  the total energy $W$ that includes  four contributions: elastic, $W^{(el)}$, magnetoelastic, $W^{(mel)}$, exchange $W^{(exc)}$, and easy-plane magnetic anisotropy, $W^{(ani)}$:
\begin{equation}\label{eq::math_en1}
W=W^{(el)}+W^{(mel)}+W^{(exc)}+W^{(ani)}
\end{equation}
where the elastic energy corresponds to isotropic elasticity 
\begin{equation}\label{eq::math_e1}
W^{(el)}=\Gamma\left[\epsilon_{11}^2+\epsilon_{22}^2+\epsilon_{33}^2+4(\epsilon_{12}^2+\epsilon_{23}^2+\epsilon_{13}^2)-2(\epsilon_{11}\epsilon_{22}+\epsilon_{22}\epsilon_{33}+\epsilon_{11}\epsilon_{33})\right] .
\end{equation}
Here $\Gamma=\mu/2$ is related with one of the Lam\'e coefficients, namely the shear modulus  $\mu$.
We keep only shear (traceless) strains as they are different in 90$^\circ$ domains and create incompatibility charges.
Similar, in magnetoelastic energy 
\begin{equation}\label{eq::math_e2}
W^{(mel)}=\frac{\Lambda}{2}\left[(\epsilon_{11}-\epsilon_{22})(n_1^2-n_2^2)+4\epsilon_{12} n_1n_2\right]
\end{equation}
we keep only nontrivial terms that couple the N\'eel vector with shear strains. Here $\Lambda=-3\mu \lambda_S$, where $\lambda_S$ is the spontaneous strain in a homogeneous state.

The exchange energy,
\begin{equation}\label{eq::math_e3}
W^{(exc)}=\frac{A}{2}|\nabla \theta|^2,
\end{equation}
and anisotropy energy, 
\begin{equation}\label{eq::math_e4}
W^{(ani)}=- \frac{1}{4}k \cos (4\theta)
\end{equation}
are  parametrised with the the exchange stiffness constant $A$  and  the anisotropy constant $k$. Anisotropy energy \eqref{eq::math_e4} corresponds to four-fold symmetry of an antiferromagnet, as was discussed above. 

Minimizing the resulting energy with respect to $u,v,w$ and $\theta$ yields
\begin{subequations}\label{eq::math_system}
\begin{equation}\label{eq::math_system1}
\Lambda \left(\cos(2\theta) \theta_y-\sin(2 \theta) \theta_x\right) +2\Gamma( u_{xx}+ u_{yy}+u_{zz})=0\end{equation}\begin{equation}\label{eq::math_system2}
\Lambda \left(\cos(2\theta)\theta_x + \sin(2\theta) \theta_y\right) +2\Gamma (v_{xx}+v_{yy}+v_{zz})=0\end{equation}
\begin{equation}\label{eq::math_system4}
w_{xx}+w_{yy}+w_{zz}=0\
\end{equation}
\begin{equation}\label{eq::math_system3}
\Lambda
\left[ (v_y-u_x) \sin(2\theta) + (u_{y}+v_{x}) \cos(2\theta)\right] -A(\theta_{xx}+\theta_{yy})+k\,\sin(4\theta)=0
\end{equation}
\end{subequations}

In order to establish the boundary conditions, we determine the spontaneous strains by minimizing setting the partial derivatives of $W$ with respect to the strain tensor to zero. The resulting straisn are

\begin{subequations}\label{eq::math_sponts}
\begin{equation}
\epsilon^{sp}_{11}=-\epsilon^{sp}\cos{2\theta} 
\label{eq::math_epsilon11}     
\end{equation}
\begin{equation}
\epsilon^{sp}_{22}=+\epsilon^{sp}\cos{2\theta} 
\label{eq::math_epsilon22}     
\end{equation}
\begin{equation}
\epsilon_{33}^{sp}=\epsilon^{sp}_{11}+\epsilon^{sp}_{22}=0 \label{epsilon33}
\end{equation}
\begin{equation}
\epsilon_{12}^{sp}=-\epsilon^{sp}\sin{2\theta} \label{epsilon12}
\end{equation}
\begin{equation}
\epsilon_{13}^{sp}=\epsilon_{23}^{sp}=0
\end{equation}
\end{subequations}

being $\epsilon^{sp}=\Lambda/\left(8\Gamma\right)$. These are the strains that each domain undergoes in the absence of stress fields which is the case for a homogeneous magnetic profile. However, in a two-domain system, they correspond to the strains observed far from the DW transition region. According to these results, all the $\epsilon_{i3}^{sp} (i=1,2,3)$ components of the spontaneous strain vanish, i.e. each AF domain is in a plane-strain state. Taking into account \eqref{eq::math_epsi} and \eqref{eq::math_sponts}, the terms $u_{zz}$, $v_{zz}$, $w_{zz}$ appearing in \eqref{eq::math_system} annihilate in each antiferromagnetic domain. Note that the same conclusion holds for the first addendum in \eqref{eq::math_system3}. Then, by assuming that the thickness of the transition layer between the two domains, i.e. the DW width (say, $x_{DW}$), is negligible with respect to the sample length ($2L$), we can safely make the assumption of disregarding the abovementioned terms. Consequently, \eqref{eq::math_system4} reduces to a bidimensional Laplace equation, whose solutions are well-known in the literature as harmonic functions and are independent on $\theta$. 

\begin{figure}[t!]
\centering \includegraphics[scale=0.5]{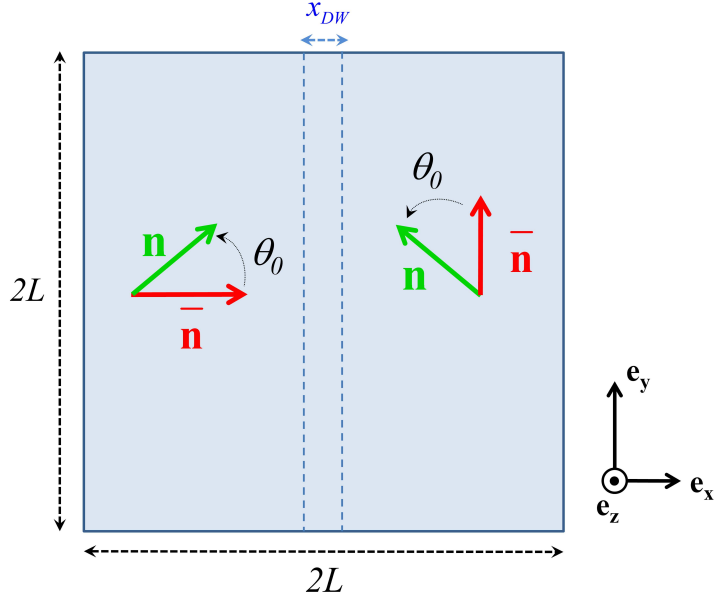}
\caption{Schematics of an antiferromagnet with 90$^\circ$ domains, together with reference axes. The N\'eel vectors here represented are: $\bar{\textbf{n}}=\left(\cos\bar{\theta},\sin\bar{\theta},0\right)$, $\textbf{n}=\left(\cos\left(\bar{\theta}+\theta_0\right),\sin\left(\bar{\theta}+\theta_0\right),0\right)$, being $\theta_0$ an arbitrary rotation angle.}
\label{fig::math_schema2}
\end{figure}

Moreover, it can be also hypothesized that any variation of the Neel vector along the $y$ axis may be ignored. so that the transition from the left domain to the right one takes place along the $x$ axis only, i.e. $\theta=\theta(x)$. Therefore, after applying the spontaneous strains \eqref{eq::math_system3} reduces to the elliptic sine-Gordon equation:

\begin{equation}\label{eq::math_cp1}
-A\,\theta''+k\sin{4\theta}=0\\
\end{equation}
where the prime denotes total derivative with respect to $x$.
Let us now look for a particular solution, indicated by $\bar{\theta}(x)$, that represents a 90$^\circ$ domain-wall where the Neel vector $\bar{\textbf{n}}=\left(\cos\bar{\theta},\sin\bar{\theta},0\right)$ points along the $x$ axis in the left domain and along the $y$ axis in the right one. Such a solution satisfies indeed the following boundary conditions
\begin{align}\label{eq::math_bc1}
\bar{\theta}(-L)=0\quad \text{and}\quad 
\bar{\theta}(+L)=\frac{\pi}{2}
\end{align}
Equation \eqref{eq::math_cp1} can be cast as 
\begin{equation}\label{eq::math_1}
\bar{\theta}'=\frac{\sin 2{\bar{\theta}}}{x_{DW}}
\end{equation}
where $x_{DW}=\sqrt{A/k}$ is representative of the domain wall width. Solution of \eqref{eq::math_1} is given by
\begin{equation}\label{eq::math_sol11}
\bar{\theta}(x)=\arctan{\left(e^{\frac{2x}{x_{DW}}}\right)}
\end{equation}

and fulfills boundary conditions \eqref{eq::math_bc1} provided that $2L\gg x_{DW}$, in line with our previous assumption. Finally, the most general scenario holds when the Neel vector is rotated everywhere counterclockwise by a constant angle $\theta_0$. In this case, it can be expressed as $\textbf{n}=\left(\cos\left(\bar{\theta}+\theta_0\right),\sin\left(\bar{\theta}+\theta_0\right),0\right)$, as schematically sketched in Fig. \ref{fig::math_schema2}. Physically such a solution corresponds to a rotation of the easy axis by $\theta_0$ from the $x$- and $y$-axis. It can be interpreted equivalently to a rotation of the DW from its assumed initial direction. Thus, the angle $\theta_0$ describes the angle between easy axes and DW orientation in the plane. 

Then, by recalling the independence of $\theta$ on $y$, the system of Poisson equations \eqref{eq::math_system1}--\eqref{eq::math_system2} reduce to
\begin{equation}\label{eq::math_2}
\begin{cases}-\Lambda \sin (2\theta)\, \theta'=-2\Gamma (u_{xx}+u_{yy})\\
+\Lambda \cos (2\theta)\,  \theta'=-2\Gamma (v_{xx}+v_{yy})
\end{cases}
\end{equation}
or equivalently, 
\begin{equation}\label{eq::math_poiseq}
\begin{cases}
-\frac{\Lambda}{x_{DW}}\left[\sin 2\theta_0\cos 2\bar{\theta}+\sin 2\bar{\theta} \cos 2\theta_0\right]\, \sin 2\bar{\theta}=-2\Gamma (u_{xx}+u_{yy})\medskip \\
+\frac{\Lambda}{x_{DW}}\left[\cos 2\theta_0\cos 2\bar{\theta}-\sin 2\bar{\theta} \sin 2\theta_0\right]\, \sin 2\bar{\theta}=-2\Gamma (v_{xx}+v_{yy})
\end{cases}
\end{equation}

Clearly, this is a system of two Poisson equations in the form $\Delta \textbf{u}= -\textbf{f}$, where the force $\mathbf{f}$ is given by the lefthandside of the eq. \ref{eq::math_poiseq}. Explicit solutions of \eqref{eq::math_poiseq} can be provided by the convolution with a  Green functions $G(x,y)$ and  read as follows:

\begin{subequations}
\begin{equation}\label{eq::math_sol3}
u(x,y)=-\frac{\Lambda}{2\Gamma x_{DW}}\iint{G(x,y,\xi,\eta) \, \sin{2\bar{\theta}}\,\sin{2(\theta_0+\bar{\theta})}\, d\xi d\eta}\end{equation}
\begin{equation}
v(x,y)=\frac{\Lambda}{2\Gamma x_{DW}}\iint{G(x,y,\xi,\eta)\, \sin{2\bar{\theta}} \,\cos{2(\theta_0+\bar{\theta})}\, d\xi d\eta}
\end{equation}
\end{subequations}

In the case of a square domain with side $2L$, the Green function can be expressed as \cite{bookgreen}:
\begin{equation}
G(x,y,\xi,\eta)=\displaystyle\frac{4}{\pi^2}\sum_{n=1}^\infty\sum_{m=1}^\infty\frac{\sin{\left(\frac{n\pi x }{2L}\right)}\sin{\left(\frac{n\pi \xi }{2L}\right)}\sin{\left(\frac{m\pi y }{2L}\right)}\sin{\left(\frac{m\pi \eta }{2L}\right)}}{n^2+m^2}
\label{eq::math_greensquare}
\end{equation}

Finally, note that eq. \eqref{eq::math_system4} reduces to the classical two-dimensional Laplace equation, whose solution is well known in the literature \cite{bookgreen}. Moreover, since it is decoupled from the previous system of Poisson equations, the displacements along $z$ axis are independent of those occurring in the $xy$ plane. 
It is shown in the appendix \ref{app::green_90} that elastic and magnetoelastic energies do not depend on the rotation angle $\theta_0$. 

On the contrary, we shall now emphasize that the incompatibility generated at the domain wall interface is responsible for volume effects proportional to $\text{tr}(\epsilon)^2=(u_x+v_y)^2$ and prove that it is not invariant under an arbitrary rotation of the Neel vector in the $xy$ plane.
Indeed, the quantity $(u_x+v_y)^2$, which reduces to $u_x^2$ under the aboev considerations, carries the dependence on $\theta_0$ according to \eqref{eq::math_ux2}. To get more insights into such a functional dependence, we compute the derivative of $u_x^2$ with respect to $\theta_0$ and, setting it to zero, we get 
\begin{align}\label{eq::math_dertet0}
\frac{\partial}{\partial \theta_0}u^2_x=0 \quad \Rightarrow \quad \left(\mathcal{P}\sin 2\theta_0+\mathcal{Q}\cos 2 \theta_0\right)\left(\mathcal{P}\cos 2 \theta_0 - \mathcal{Q}\sin 2\theta_0\right)=0
\end{align} 
where
\begin{equation}
\mathcal{P}=\iint{G(x,y,\xi,\eta)\sin{2\bar{\theta}}\cos(4\bar{\theta})\, d\xi d\eta}, \quad
\mathcal{Q}=\iint{G(x,y,\xi,\eta)\sin{2\bar{\theta}}\sin(4\bar{\theta})\, d\xi d\eta}
\end{equation}
Then, \eqref{eq::math_dertet0} has two zeros:
\begin{equation}
\theta^{(1)}_0=\frac{1}{2}\arctan{\left(\frac{\mathcal{P}}{\mathcal{Q}}\right)}, \qquad \theta^{(2)}_0=\frac{1}{2}\arctan{\left(-\frac{\mathcal{Q}}{\mathcal{P}}\right)}
\end{equation}

By using \eqref{eq::math_greensquare}, we deduce that $\mathcal{P}$ vanishes whereas $\mathcal{Q}$ is a non-zero quantity. The critical points are: $\theta^{(1)}_0=0, \pi$, that are maxima, and $\theta_0^{(2)}=\pm \pi/4$, that are minima. This result suggests that, despite the governing equations do not account for volume effects, they arise from the incompatibility between spontaneous strains at the domain wall interface. Moreover, we found that the DW configuration characterized by a $\theta_0=\pi/4$ rotation constitutes the optimal setup where volume effects are minimized. 

In the following we validate the mentioned theoretical predictions.


\section{Numerical results}
\label{sec::numerics}

To validate the previous findings, we integrate numerically the system of Poisson equations \eqref{eq::math_poiseq} by means of COMSOL Multiphysics$^{(R)}$ \cite{COMSOL}.
To this aim, we consider the set of parameters reported in Table \ref{Tablepar1} from which we deduce the derived quantities reported in Table \ref{Tablepar2}.

\begin{table}[htbp]
\centering
\begin{tabular}{|c|c|}
\hline 
Parameter& Value\\
\hline
$\lambda$&$9.5\times 10^{10} \, J/m^3$\\
$\mu$&$6.5\times 10^{10} \, J/m^3$\\
$\lambda_S$&$-100\times 10^{-3}$\\
$2L$&$500 \, nm$\\
$A$&$1\times 10^{-12}\, J/m$\\
$k$&$1\times 10^{3}\, J/m^3$\\
\hline
\end{tabular} 
\\
\caption{Parameters used in the numerical simulation.}
\label{Tablepar1}
\end{table}
\begin{table}[htbp]
\centering
\begin{tabular}{|c|c|c|}
\hline 
Parameter& Expression&Value\\
\hline
$x_{DW}$&$\sqrt{A/k}$&$31.6 nm$\\
$\Gamma	$&$\mu/2$&$3.25\times 10^{10} J/m^3$\\
$\Lambda$&$-3\mu\lambda_S$&$1.95\times 10^{10} J/m^3$\\	
$\epsilon^{sp}$ & $\Lambda/(8\Gamma)$&$0.075$\\
\hline
\end{tabular}
\\
\caption{Derived parameters.}
\label{Tablepar2}
\end{table}

The previous system is integrated over a computational square domain of side $2L=500 \,nm$ which, according to the above parameter set, is much larger than the domain wall width $x_{DW}=31.6 \,nm$, in line with the assumption made in Section \ref{sec::math}. 
We also used \eqref{eq::math_sponts} as boundary conditions at the vertical edges $x=\pm L$. On the other hand, at the horizontal edges $y=\pm L$, having normal $\hat{\textbf{q}}=\pm \textbf{e}_y$, we implemented the following piecewise conditions:

\begin{equation}
\left(\hat{\textbf{q}} \cdot \nabla\right) u (x) = 
\begin{cases}
-\epsilon^{sp} \sin\left(2 \theta_0\right) \phantom{spaceeeeeeeeeeeeeeeee.} \text{for}\quad  x \in \left[-L;-L+\bar{L}\right] \medskip \\
\displaystyle -\epsilon^{sp}\frac{\sin\left(2 \left(\theta_0+\pi/2\right)\right)-\sin\left(2 \theta_0\right)}{2\left(L-\bar{L}\right)}x  \quad \text{for}\quad x \in \left[-L+\bar{L};L-\bar{L}\right] \medskip \\
-\epsilon^{sp} \sin\left(2 \left(\theta_0 + \pi/2\right)\right) \phantom{spaceeeeeeeee.} \text{for}\quad x \in \left[L-\bar{L};L\right] \\
\end{cases}
\end{equation}
 and 
\begin{equation}
\left(\hat{\textbf{q}} \cdot \nabla\right) v(x) = 
\begin{cases}
\epsilon^{sp} \cos\left(2 \theta_0\right) \phantom{spaceeeeeeeeeeeeeeeeeee.} \text{for}\quad  x \in \left[-L;-L+\bar{L}\right] \medskip \\
\displaystyle \epsilon^{sp}\frac{\cos\left(2 \left(\theta_0+\pi/2\right)\right)-\cos\left(2 \theta_0\right)}{2\left(L-\bar{L}\right)}x \phantom{-}  \quad \text{for}\quad x \in \left[-L+\bar{L};L-\bar{L}\right] \medskip \\
\epsilon^{sp} \cos\left(2 \left(\theta_0 + \pi/2\right)\right) \phantom{spaceeeeeeeeeee.} \text{for}\quad x \in \left[L-\bar{L};L\right] \\
\end{cases}
\end{equation}

Here, the parameter $\bar{L}$ is set to $100 \,nm$ and the rotation angle $\theta_0$ is varied in the range $[0,\pi/2]$.

\begin{figure}[b!]
\centering \includegraphics[scale=0.8]{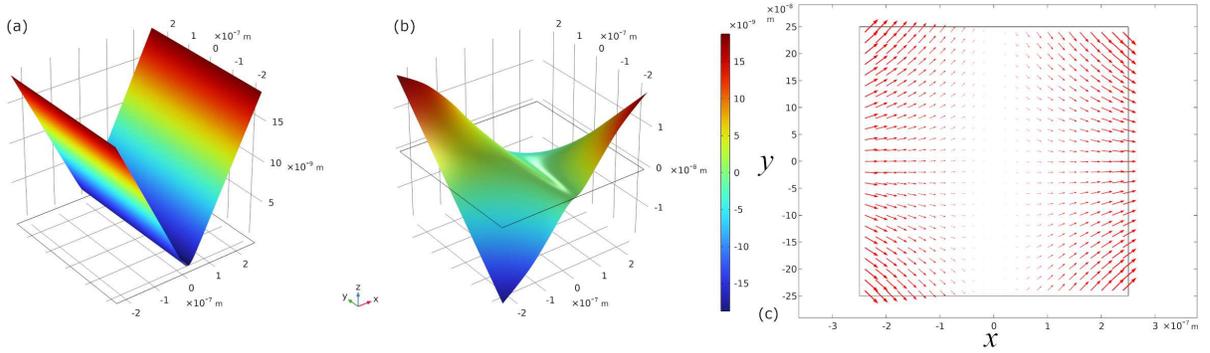}
\caption{(a,b) The $u$ and $v$ components of the displacement vector and (c) its vector representation obtained for $\theta_0 = 0$.}
\label{0_90}
\end{figure}

\begin{figure}[b!]
\centering \includegraphics[scale=0.8]{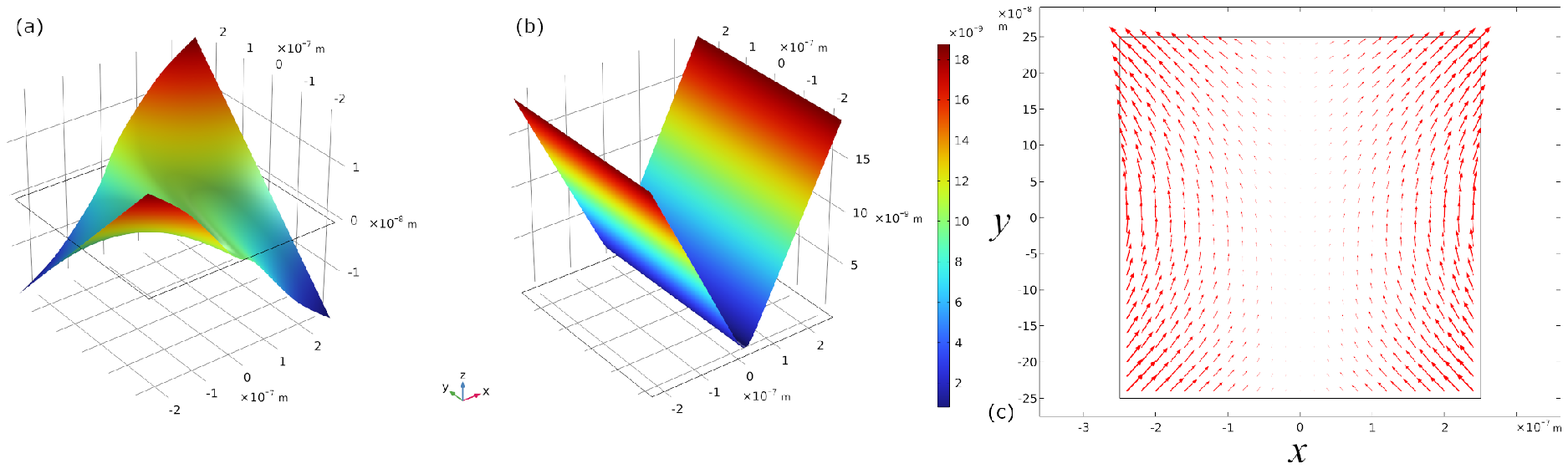}
\caption{(a,b) The $u$ and $v$ components of the displacement vector and (c) its vector representation obtained for $\theta_0 = \pi/4$.}
\label{45_135}
\end{figure}

As illustrative examples, in Figs.\ref{0_90},\ref{45_135} we show the numerically-obtained solutions for $u(x,y)$ and $v(x,y)$ for $\theta_0=0$ and $\theta_0=\pi/4$, respectively. In particular, the panels (c) point out that the overall displacements are minimal at the DW interface, $x=0$. Also, since the material tends to elongate (compress) along the direction orthogonal (longitudinal) to $\textbf{n}$, different deformations are induced as a function of the rotation angle $\theta_0$.

\begin{figure}[t!]
\centering
\includegraphics[scale=0.8]{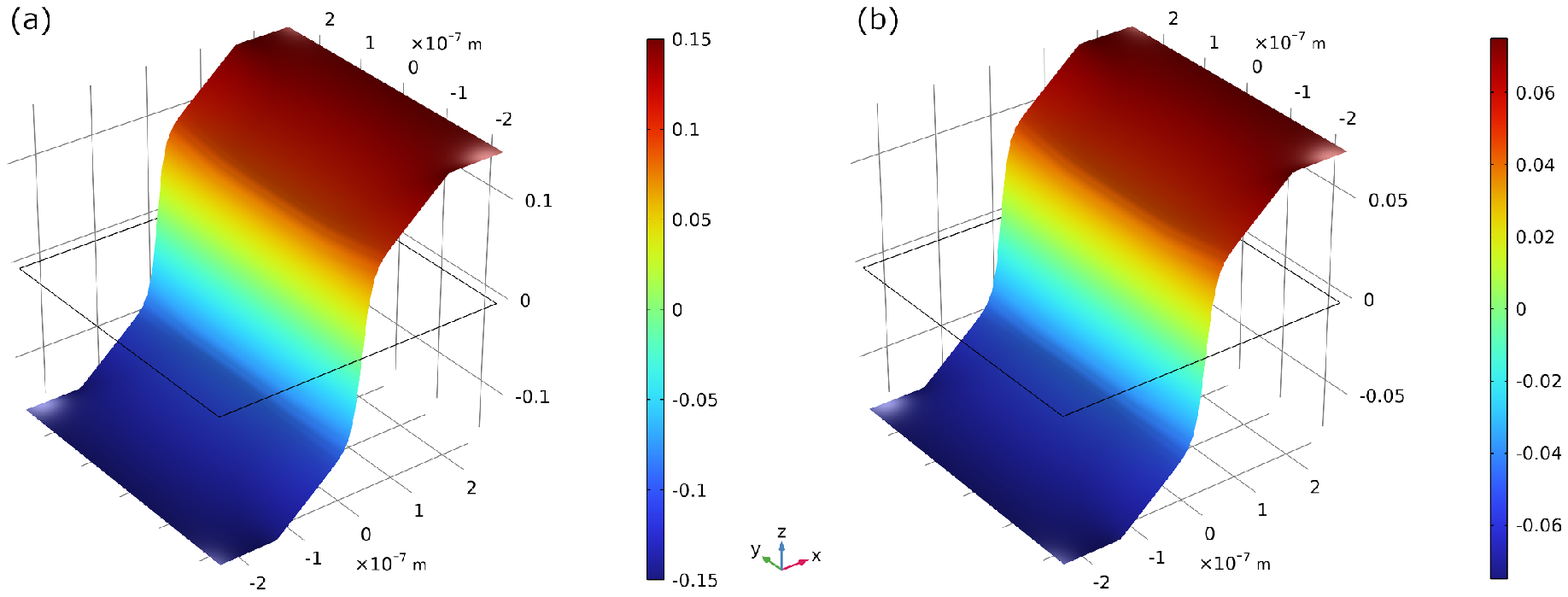}
\caption{Spatial dependence of (a) $\left(\epsilon_{11}-\epsilon_{22}\right)(x,y)$ obtained for $\theta_0=0$ and (b) $\epsilon_{12}(x,y)$ for $\theta_0=\pi/4$.}
\label{figepsspont}
\end{figure}

Analogously, in Fig. \ref{figepsspont} we represent the spatial dependence of $\epsilon_{11}-\epsilon_{22}$ for $\theta_0=0$ and $\epsilon_{12}$ for $\theta_0=\pi/4$. As it can be noticed, according to the symmetry of the governing equations \eqref{eq::math_poiseq} and boundary conditions \eqref{eq::math_sponts}, the role of the two components of  displacement vector as well as that of normal and shear strains is exchanged in the two configurations. Moreover, such figures reveal that a smooth transition takes place along the $x$ axis between the spontaneous strains at the lateral edges of the sample, $x=\pm L$.


\begin{figure}[t!]
\centering
\includegraphics[scale=0.8]{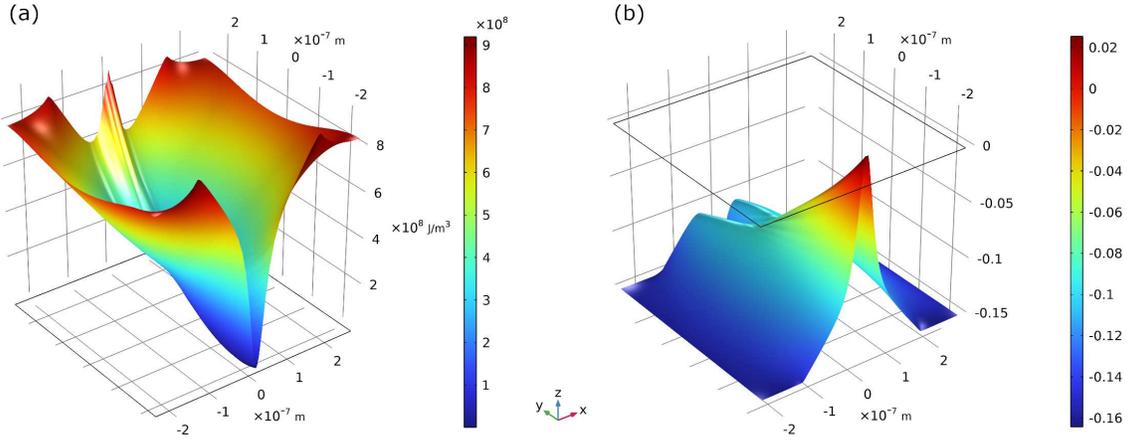}
\caption{Spatial dependence of (a) elastic and (b) magnetoelastic energies.}
\label{energies}
\end{figure}

Then, let us consider the elastic $W^{(el)}$ and magnetoelastic $W^{(mel)}$ energy contributions. It has been checked numerically that they are unaffected by variations of $\theta_0$, as proved theoretically. In Fig.\ref{energies} we depict their spatial distributions.

\begin{figure}[htbp]
\centering
\includegraphics[scale=0.5]{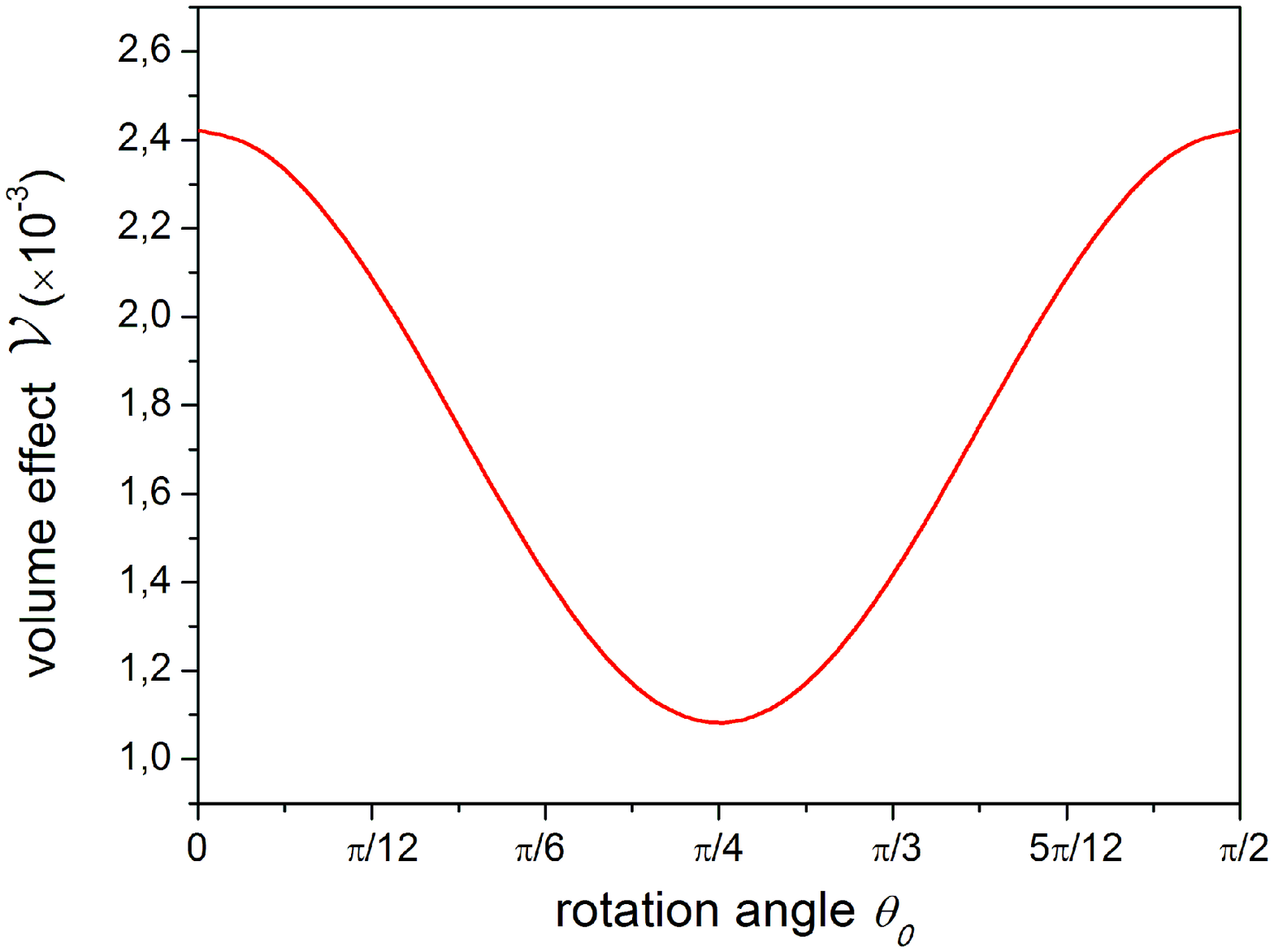}
\caption{Volume effect $\mathcal{V} $ as a function of the rotation angle $\theta_0$. }
\label{volu_rot}
\end{figure}

\newpage
Finally, we computed numerically the variation of the quantity representative of volume effects
\begin{equation}
\mathcal{V}=\int_\Omega{(u_x+v_y)^2\, dxdy}
\end{equation}
as a function of $\theta_0$. Here $\Omega=[-L/2,L/2]\times[-L/2,L/2]$ is a square subset of the whole computational domain that has been considered to avoid spurious boundary effects. Numerical results shown in Fig.\ref{volu_rot} point out that this quantity does vary with the rotation angle and exhibits a cosinusoidal behavior with a minimum at $\theta_0=\pi/4$, in line with theoretical expectations.

\section{Conclusions}
\label{sec::conclusions}
We calculated strain fields and  corresponding energy of the domain wall depending on orientation of the domain wall plane with respect to easy magnetic axes. Our calculations show angular dependence with 90$^\circ$ periodicity, consistent with the general symmetry properties of the strain tensor.  Energy minima, and, hence, favourable orientations of the domain wall, correspond to hard magnetic directions in which incompatibility between spontaneous strains in both domains vanishes (so called Nye walls, according to \cite{Kleman1972}). Anisotropy of the domain wall energy is due to magnetoelastic effects. Moreover, surprisingly, magnetoelastic contribution originated from the volume additional strains, though spontaneous strains include only shear (traceless) components. This is an illustration of gradient elasticity, where the strain components with different symmetry properties are coupled though the strain gradient.  Similar, thermomagnetoelastic  effect  was recently observed in Ref. \cite{Meer2021}: temperature-induced gradients of isotropic strains created shear strains and via magnetoelastic mechanism induced reorientation of the N\'eel vector in the sample. In our case the gradients of the shear strain produced by the domain wall induce isotropic strains. These isotropic strains can be potentially manipulated by heating or cooling of the sample, which opens a way for  control the domain wall distributions within an antiferromagnet.
 
BK and OG
acknowledge that this work was funded by the Deutsche Forschungsgemeinschaft (DFG, German Research Foundation), TRR 173-268565370 (Project Nos. B12).  O.G. acknowledges  support by the Deutsche Forschungsgemeinschaft  TRR 288-422213477 (project A09) and the EU FET Open RIA Grant No. 766566.

GC and PV acknowledge the financial support from INdAM-GNFM and MUR (Italian Ministry of University and Research) through project PRIN2017 n. 2017YBKNCE entitled \textquotedblleft Multiscale phenomena in Continuum
Mechanics: singular limits, off-equilibrium and transitions\textquotedblright. PV also acknowledges the research project \textquotedblleft Mathematical Methods in Non Linear Physics
(MMNLP)\textquotedblright by the Commissione Scientifica Nazionale -- Gruppo 4 -- Fisica Teorica of the Istituto Nazionale di Fisica Nucleare (INFN).
 

\bibliography{Ref}
\bibliographystyle{unsrtdin}

\end{document}